\documentclass[twocolumn,showpacs,preprintnumbers,amsmath,amssymb]{revtex4}
\usepackage{graphicx}
\usepackage{epsf}
\usepackage{dcolumn}
\usepackage{bm}

\newcommand\f[2]{\frac{#1}{#2}}

\def\beq{\begin{equation}} 
\def\eeq{\end{equation}} 
\def\beeq{\begin{eqnarray}} 
\def\eeeq{\end{eqnarray}} 
 
\def\to{\rightarrow} 
\def\nn{\nonumber}

\begin{document}

\title{~~~~\\ Nuclear parton distributions at next to leading order.\\ ~~~~~~~ }
\author{\large   D. de Florian, R. Sassot\footnote{Partially supported by Conicet, ANPCyT, UBACyT and Fundaci\'on Antorchas} }  
\email{deflo@df.uba.ar, sassot@df.uba.ar}

\affiliation{
 Departamento de F\'\i sica, Facultad de Ciencias Exactas y Naturales,
 Universidad de Buenos Aires, Pabell\'on I, Ciudad Universitaria 
                       (1428) Buenos Aires, Argentina }
\date{\today }

\begin{abstract}
We perform a next to leading order QCD global analysis of nuclear deep 
inelastic scattering and Drell-Yan data using the convolution approach 
to parameterize nuclear parton densities. We find both a significant 
improvement in the agreement with data compared to previous extractions,
and substantial differences in the scale dependence of nuclear effects
compared to leading order analyses.
\end{abstract}

\maketitle

\section{Introduction}
\label{sec:introito}
Ever since the discovery that quark and gluons in bound nucleons show 
momentum distributions noticeably different to those measured in free
or less bound nucleons \cite{EMC}, two decades ago, the precise 
determination 
of nuclear parton densities has awakened growing attention, driving 
both increasingly precise and comprehensive nuclear structure functions
measurements \cite{arneodo}, and more refined theoretical understanding of the underlying physics.

The precise knowledge of nuclear parton distribution functions (nPDF) is 
not only required for a deeper understanding of the mechanisms associated 
to nuclear binding from a QCD improved parton model perspective, but it is
also the starting point for the analyses of a wide variety of future and
ongoing high energy physics experiments, such as heavy ion collisions at 
RHIC \cite{RHIC}, proton-nucleus collisions to be performed at LHC \cite{LHC},
or neutrino-nucleus interactions in long baseline neutrino experiments 
\cite{neutrino}, for example. Consequently,
the accuracy of nPDF is rapidly evolving into a key issue in many areas of 
particle physics.

From the point of view of perturbative QCD, the extraction of nPDF can be 
done in close analogy to what is done for free nucleons: they are considered
as non perturbative inputs, to be inferred from data, whose relation to the
measured observables and energy scale dependence are computed order by order 
in perturbation theory. Although one can not discard larger higher twist 
power corrections than in the case of free nucleons, or even some sort of
nuclear recombination effect, factorization and universality of nPDF are 
expected to hold in a very good approximation, over a wide kinematical range 
in the present experiments.   

At variance with nucleon PDF, which driven by the demand of 
more and more precise predictions, have attained in the last years an 
impressive degree of accuracy and refinement, nPDF extractions are in a 
considerably earlier stage of the development. Not only the number and 
diversity of nuclear data is much more reduced, but the analyses are 
restricted to leading order accuracy (LO), with rather crude parameterizations 
of nuclear effects which lead to poor $\chi^2/d.o.f$ values in global QCD 
fits to data \cite{EKS98,HKM}.                 

There are also some caveats inherent to the particular approaches implemented 
so far, that define nPDF in terms of nucleon PDF and a multiplicative nuclear 
correction factor at a given initial energy scale, from where they are 
evolved. 
In the first place, nPDF defined in this way have their momentum fraction 
per nucleon restricted to be less or equal than unity, what excludes from
the description, and also from the evolution, a portion of its natural 
kinematical range. Of course, unless one is specifically interested in that 
region, this approximation is expected to imply a minor correction, although
one can not discard the excluded region as a source of new and interesting
information.  

Much more problematic in these  approaches is that 
the actual shape of the nuclear correction factor required to reproduce 
accurately the data implies very capricious functions, with a large number 
of parameters, and which in practice precludes the numerical computation of 
the scale dependence at next to leading order accuracy (NLO). The actual 
computation of structure functions and evolution equations at this order 
implies several convolution integrals very difficult to evaluate unless Mellin 
transform techniques are applied. 

In this paper we show that a much more convenient alternative to deal with 
nuclear effects is to define
nPDF using a convolution approach. In such framework the free nucleon parton 
densities are convoluted with very simple weight functions that parameterize 
nuclear effects. The convolution method naturally takes into account the 
actual range of nucleon momentum fractions, allows via Mellin transform 
techniques a straightforward numerical evaluation of the NLO scale dependence,
lead to extraordinarily accurate nPDF with relatively few parameters, and 
finally, allow to interpret the nuclear modifications in terms of a very 
simple mechanism of rebalance of momentum fractions between the distributions.
The success of the convolution approach comes from the fact that the momentum 
fraction dependence of nuclear effects is strongly correlated to that of 
partons in free nucleons, as shown in re-scaling models \cite{CGC}, 
feature which is explicitly included by the convolution.

As an example of feasibility of the convolution approach we obtain for the 
first time a full NLO extraction of nPDF from a large number of nuclear DIS 
\cite{NMC1,SLAC1,NMC2,NMC3} and Drell Yan data \cite{772}. We also asses the 
differences between the LO and NLO extractions, finding that although the 
quality of fits to present data is comparable in both approximations, there 
are important differences between the $Q^2$ dependence of the nuclear 
correction factors using either LO or NLO extractions. This, for example, 
questions  the use of LO factors with NLO nucleon parton densities to generate
 nPDF in NLO computations, as it is common practice. 

In the following section we summarize the main motivation and features of the 
convolution approach. In the third section we present the details and outcome
of the NLO and LO nPDF extractions, comparing our results with previous LO 
analyses. In the last section we discuss the differences between LO and NLO
extractions of nPDF, computing the one hadron production cross section with 
them as and example, and present our conclusions.
       
\section{nPDF}
\label{sec:npdf}

The description of DIS processes off nuclear targets, $e A \to e' X$, is customarily done in terms of the hard scale $Q^2$, defined as minus the virtuality of
the exchanged photon, and a scaling variable $x_A$, analogue to the Bjorken variable used in DIS off nucleons,
\begin{equation}
\label{eq:f0}
Q^2\equiv -q^2,\,\,\,\,\,\,\,\,\,\,\,\,\,\,\,\,\,\, x_A \equiv \frac{Q^2}{2\, p_A\cdot q},
\end{equation}
respectively. Here $p_A$ is the target nucleus momentum and consequently, $x_A$ is kinematically restricted to $0<x_A<1$, just as the standard Bjorken variable. Alternatively, one can define another scaling variable $x_N\equiv A\,x_A$, where $A$ is the mass number of the nucleus.  Under the assumption that the nucleus momentum $p_A$ is evenly distributed between that of the constituent nucleons
$p_N=p_A/A$, this variable resembles the Bjorken variable corresponding 
to the scattering off free nucleons, $x_N\equiv Q^2/(2p_N\cdot q)$. However
in nuclear scattering context of course it spans the interval $0<x_N<A$, 
by definition, and reflecting the fact that a parton may in principle carry more
than the average nucleon momentum.

When discussing nPDF the usual approach is to propose a very simple relation 
between the parton distribution of a {\it proton} bound in the nucleus, $f_i^A$, and those for free protons $f_i$,
\begin{equation}
\label{eq:f1}
f_i^A(x_N,Q_0^2) = R_i(x_N,Q_0^2,A,Z)\, f_i(x_N,Q_0^2),
\end{equation}
in terms of a multiplicative nuclear correction factor $R_i(x_N,Q^2,A,Z)$, specific
for a given nucleus $(A,Z)$, parton flavor $i$, and initial energy scale 
$Q^2_0$. This description is convenient since the ratio $R_i(x_N,Q^2,A,Z)$ 
compares directly the parton densities with and without nuclear effects, and
is closely related to the most usual nuclear DIS observables, which are the 
ratios between the nuclear and deuterium structure functions. Indeed, we will 
use this picture  to simplify the presentation of the output of our analysis.
However, this is not
the best suited way to parameterize the effects in the intermediate steps of 
the analysis, nor the best alternative for higher order numerical computations,
as it was mentioned previously.    

Since $f_i(x_N,Q^2)$ are defined for $0<x_N<1$, nPDF defined as in Eq.(\ref{eq:f1}) are a priori restricted to that same range at the initial scale. 
In most analyses the evolution equations are also constrained to $0<x_N<1$, 
loosing the possibility of analyzing nuclear effects beyond $x_N>1$ at any 
other scale. Notice that even if nPDF were constrained to be zero for $x_N>1$ 
at a given scale, the complete evolution equations would produce non-zero 
values for other scales in the range of $x_N$ neglected by the approach, 
leading to a possible violation of momentum conservation.
From another side, the direct parameterization of the ratios
$R_i(x_N,Q^2,A,Z)$ require many parameters and even $x_N$-dependencies whose 
Mellin moments can not be put into a closed expression.

A much more adequate alternative to Eq.(\ref{eq:f1}) is to relate nPDF to standard PDF by means of a convolution
\begin{equation}
\label{eq:f2}
f_i^A(x_N,Q_0^2) = \int_{x_N}^A\,\,  \f{dy}{y} W_i(y,A,Z)\,\, f_i(\f{x_N}{y},Q_0^2).
\end{equation}
where the weight function $W_i(y,A,Z)$ parameterize now the nuclear effects and can be thought as the effective density of nucleons within the nucleus carrying a fraction $y/A$ of its longitudinal momentum. Besides allowing the full kinematical range for nPDF, this kind of approach has been shown to lead to remarkably good parameterizations of nuclear effects with very few parameters, and with a smooth $A$-dependence \cite{Z}. 
For example, neglecting nuclear effects, the effective nucleon density is just $W_i(y,A,Z)= A \,\delta(1-y)$, and simple shifts in the momentum fraction carried by nucleons $W_i(y,A,Z)= A \,\delta(1-y-\epsilon)$ have been shown to reproduce in a very good approximation many features of the known nuclear modifications to structure functions.

The convolution approach also allows to perform the evolution in Mellin space, 
which is much more convenient numerically, and almost mandatory for NLO 
accuracy. Defining $y_A \equiv 
\f{y}{A}$  Eq.(\ref{eq:f2}) reads
\begin{equation}
\label{eq:f3}
f_i^A(A\,x_A,Q_0^2) = \int_{x_A}^1\,\,  \f{dy_A}{y_A}\,\, W_i(A y_A,A,Z)\, f_i(\f{x_A}{y_A},Q_0^2)  \,\,\,\,\,\,\, \,\,\,\,\,\,\,\,\,\,\,\,\,\, \,\,\,\,\,\,\,
\end{equation}
and in Mellin space, i.e. taking moments in both sides of Eq.(\ref{eq:f3}),
\begin{equation}
\label{eq:fn}
\hat{f}_i^A(N,Q_0^2) =  \hat{W}_i(N,A,Z)\, \hat{f}_i(N,Q_0^2)
\end{equation}
where 
\begin{equation}
\label{eq:ffn}
\hat{f}_i^A(N,Q_0^2)\equiv \int_0^1 dx\, x^{N-1} f_i^A(A\,x,Q_0^2)
\end{equation}
and similarly for $\hat{f}_i(N,Q_0^2)$ with $A=1$, and
\begin{equation}
\label{eq:fwn}
\hat{W}_i(N,A,Z)\equiv \int_0^1 dx\, x^{N-1}  W_i(A\, x,A,Z).
\end{equation}
Notice that the moments are defined in terms of the correct scaling variable 
$x_A$, and that the evolution of the nPDF moments $\hat{f}_i^A(N,Q_0^2)$ can 
be managed with a standard evolution code in Mellin space.

Nuclear structure functions are defined as the average of the proper 
combination of the bound proton and neutron structure functions as 
\begin{equation}
A\, F^{A}_2(x,Q^2)= Z\, F^{p/A}_2(x,Q^2) + (A-Z)\, F^{n/A}_2(x,Q^2),
\end{equation}
where both bound nucleon structure functions can be written in terms of the 
corresponding nuclear parton distributions in Mellin space as
\begin{align}
\label{eq:ff2}
 F^{p/A}_2&(N-1,Q^2)= \sum_{q,\bar{q}} e_q^2 \left\{ \hat{f}_i^A(N,Q^2) \right. \\ 
&\left. + \f{\alpha_s}{2\pi} \left[ C_q^{(1)}(N) \,  \hat{f}_i^A(N,Q^2) +  C_g^{(1)}(N) \,  \hat{f}_g^A(N,Q^2) \right] \right\}  \nn \, .
\end{align}
The first term in the right hand side of Eq.(\ref{eq:ff2}) corresponds to the 
leading-order contribution while the second represents the next-to-leading 
order correction. Expressions for the NLO quark and gluon coefficients $ C_q^{(1)}(N)$ and $ C_g^{(1)}(N)$ can be found in \cite{nlocoeff}. In our analysis we consider only 3 active flavors and neglect the contribution from heavy quarks to the structure function.

Considering that DIS data does not allow to determine all combinations of flavors, and that nuclear effects may be expected to be isospin invariant in a first approximation, it seems to be reasonable  to introduce only three independent $W_i(y,A,Z)$: one for the valence distributions, another for the light sea, and the last for the gluons. In this way
\begin{eqnarray}
u_v^A(N,Q_0^2) &=&   \hat{W}_v(N,A,Z) \,\,  u_v(N,Q_0^2) \nn  \\
d_v^A(N,Q_0^2) &=&   \hat{W}_v(N,A,Z)  \,\,  d_v(N,Q_0^2) \nn  \\
\bar{u}^A(N,Q_0^2) &=&   \hat{W}_s(N,A,Z)  \,\,  \bar{u}(N,Q_0^2) \nn  \\
\bar{d}^A(N,Q_0^2) &=&   \hat{W}_s(N,A,Z)  \,\,  \bar{d}(N,Q_0^2) \nn  \\
g^A(N,Q_0^2) &=&   \hat{W}_g(N,A,Z)  \,\,  g(N,Q_0^2) 
\end{eqnarray}
Notice that in many proton parton densities, and in particular in GRV98 
\cite{GRV98}, the one will be used in the following, the strange and heavy quark densities vanish 
at the low initial scale and there is no need to introduce additional weight 
functions for the latter. The parton distributions for the bound neutron can be obtained from the ones above by isospin symmetry.

An interesting feature in this approach is that since non-singlet 
combinations of parton densities, as those for valence quarks, evolve 
independently, and both  $\hat{f}_i^A(N,Q^2)$ and $\hat{f}_i(N,Q^2)$
obey the same evolution equations, 
 the moments of the weight function $\hat{W}_i(N,A,Z)$ are scale independent \footnote{Notice that this is not the case for the ratios $R$ defined in $x-$space as in Eq.(\ref{eq:f1})}.
 This is not the case for sea or gluon weight functions, since they are
 affected by singlet evolution resulting in a scale dependence due to quark and gluon mixing and therefore needed to be defined at a particular scale.

Charge, baryon number and momentum conservation imply constraints to be 
satisfied by the weight functions:
\begin{align}
\label{eq:r1}
  \hat{W}_v(N=1,A,Z) &= 1  \\
\label{eq:r2}
u_v^A(2,Q_0^2) + d_v^A(2,Q_0^2)&   \nonumber \\
+ 2 \bar{u}^A(2,Q_0^2) + 2 \bar{d}^A(2,Q_0^2) 
+  g^A(2,Q_0^2) &=1 
\end{align}

The best results are obtained using for the valence distributions weight 
functions like
\begin{eqnarray}
\label{eq:rv}
W_v(y,A,Z) =\hspace*{60mm}\\
 A\,\left[\, a_v\, \delta(1-\epsilon_v -y)+(1-a_v)\, \delta(1-\epsilon_{v'} -y) \right]\hspace*{4mm}\nonumber \\
\hspace*{10mm} +n_v \left(\f{y}{A}\right)^{\alpha_v} \left(1-\f{y}{A}\right)^{\beta_v}
 +n_s \left(\f{y}{A}\right)^{\alpha_s} \left(1-\f{y}{A}\right)^{\beta_s}
\nonumber
\end{eqnarray}
where the first two terms may be interpreted as reduction in the parent 
nucleon longitudinal momentum fraction, and in spite of their simplicity 
reproduce accurately EMC and Fermi motion effects. Indeed, with just these 
three parameters the fit reproduces fairly well the large $x_N$ data, however 
it fails to account for antishadowing effects at intermediate $x_N$, where 
valence distributions still dominate. 

In order to include antishadowing effects we add the third term in Eq.
(\ref{eq:rv}), which  induces a small enhancement of the distributions with 
$n_v>0$ and a mild $x_N$ dependence given by the parameters $\alpha_v$ and 
$\beta_v$. Notice that the convolution integral dilutes the $x_N$ dependence 
producing an effect similar to those predicted by parton recombination models 
\cite{Close}.
It would be pointless to try to extract from the data the nuclear modification
to valence distributions in the low $x_N$ region, since this region is clearly
dominated by sea and gluon densities, however the first three terms in 
Eq.(\ref{eq:rv}) violate charge conservation. In order to remedy this situation
we include the forth term, similar to the third but where charge conservation
forces $n_s<0$. In this case the weight function can not be interpreted as a 
probability density but as the result of the mechanism that compensate re-scaling and recombination effects. The 
parameters $\alpha_s$ and $\beta_s$ are taken to be the parameters used for 
the sea quark densities, fixed by the low $x_N$ behavior of the data, while 
both $n_v$ and $n_s$ are fixed by momentum and charge conservation, 
Eqs.(\ref{eq:r1}) and (\ref{eq:r2}).

For sea quarks and gluons, the argument is just the opposite: the fit to data is not sensitive to any nuclear modification at high or intermediate $x_N$, so the best choice for sea weight functions is found to be
\begin{eqnarray}
W_s(y,A,Z) = A\,\delta(1-y)+ \frac{a_s}{N_s} \left(\f{y}{A}\right)^{\alpha_s} \left(1-\f{y}{A}\right)^{\beta_s}
\end{eqnarray}
where the first term gives the gluon distribution in free protons and the 
eulerian function (specifically the parameter $\alpha_s$) drives the low 
$x_N$ shadowing ($a_s$ is found to be negative). The three parameters, $a_s$,
$\alpha_s$ and $\beta_s$ are strongly constrained by the structure functions 
ratios at small $x_N$ and the Drell-Yan ratios. $N_s$ is just the 
normalization of the eulerian function, $N_s= B(\alpha_s+2,\beta_s+1)$. 
Similarly, for gluons
\begin{eqnarray}
W_g(y,A,Z) = A\,\delta(1-y)+\frac{a_g}{N_g} \left(\f{y}{A}\right)^{\alpha_g} \left(1-\f{y}{A}\right)^{\beta_g}.
\end{eqnarray}
Here, $a_g$  is strongly constrained by the $Q^2$-dependent $F_2^{Sn}/F_2^{C}$
 data, whereas
the exponents are taken to be the same as for the sea distributions and $N_g=N_s$. No significant improvement is found assigning independent parameters for them.
Since $n_v$ and $n_s$ are constrained by momentum conservation in Eqs.(\ref{eq:r1}) and (\ref{eq:r2}), there are 9 independent parameters $\epsilon_v, \epsilon_{v'}, a_v, \alpha_v, \beta_v a_s, \alpha_s, \beta_s,$ and $ a_g$ for each 
nucleus. Since no data is available on different isotopes of the same nucleus,
in the following we drop the dependence on $Z$. The $A$ dependence of all the parameters can be written 
\begin{equation}
\epsilon_i= \gamma_i + \lambda_i A^{\delta_i} 
\end{equation}
The full parameterization of both the $A$ and $x_N$ dependence of nuclear effects imply in principle quite a lot of parameters (27), however 
the mild $A$ dependence found in some of them  allow to reduce their number. 
For example, $\alpha_s$, $\alpha_v$ and $a_v$ can be taken as constant in $A$
loosing just a few units in $\chi ^2$ but eliminating 6 parameters.

\section{Results}
\label{sec:res}

In the following section we present results from the LO and NLO fits to 
nuclear data using the convolution approach. The data analysed include the 
most recent NMC and SLAC-E139 DIS structure functions ratios to deuterium 
and carbon, selecting those measurements corresponding to  $Q>1$ GeV,
and also E772 Drell-Yan dilepton cross sections from proton 
nucleus collisions, as listed in Table 1, rendering a total number of 420 data points.

\begin{table}
\caption{\label{tab:table1} Nuclear data included in the fit. }
\begin{ruledtabular}
\begin{tabular}{llcc} 
Measurement              & Collaboration    &   Refs.       & \# data \\ \hline
$F_2^{He}/F_2^{D}$ & NMC       &  \cite{NMC1}  & 18        \\ 
                  & SLAC-E139 &  \cite{SLAC1} & 18        \\ 
$F_2^{Be}/F_2^{D}$ & SLAC-E139 &  \cite{SLAC1} & 17        \\ 
$F_2^{C}/F_2^{D}$ & NMC       &  \cite{NMC1}  & 18        \\ 
                  & SLAC-E139 &  \cite{SLAC1} & 7        \\ 
$F_2^{Al}/F_2^{D}$ & SLAC-E139 &  \cite{SLAC1} & 17        \\ 
$F_2^{Ca}/F_2^{D}$ & NMC       &  \cite{NMC1}  & 18        \\ 
                  & SLAC-E139 &  \cite{SLAC1} & 7        \\
$F_2^{Fe}/F_2^{D}$ & SLAC-E139 &  \cite{SLAC1} & 23       \\  
$F_2^{Ag}/F_2^{D}$ & SLAC-E139 &  \cite{SLAC1} & 7       \\  
$F_2^{Au}/F_2^{D}$ & SLAC-E139 &  \cite{SLAC1} & 18       \\
$F_2^{Be}/F_2^{C}$ & NMC       &  \cite{NMC2}  & 15        \\   
$F_2^{Al}/F_2^{C}$ & NMC       &  \cite{NMC2}  & 15        \\   
$F_2^{Ca}/F_2^{C}$ & NMC       &  \cite{NMC2}  & 15        \\   
$F_2^{Fe}/F_2^{C}$ & NMC       &  \cite{NMC2}  & 15        \\   
$F_2^{Pb}/F_2^{C}$ & NMC       &  \cite{NMC2}  & 15        \\   
$F_2^{Sn}/F_2^{C}$ & NMC       &  \cite{NMC3}  & 145        \\   
$\sigma^{C}_{DY}/\sigma^{D}_{DY}$  & E772 & \cite{772} & 9 \\
$\sigma^{Ca}_{DY}/\sigma^{D}_{DY}$ & E772 & \cite{772} & 9 \\
$\sigma^{Fe}_{DY}/\sigma^{D}_{DY}$ & E772 & \cite{772} & 9 \\
$\sigma^{W}_{DY}/\sigma^{D}_{DY}$  & E772 & \cite{772} & 9 \\
\hline
Total              &           &               & 420 \\
\end{tabular}
\end{ruledtabular}
\end{table}

\setlength{\unitlength}{1.mm}
\begin{figure}[t]
\begin{picture}(100,87)(0,0)
\put(7,-17){\mbox{\epsfxsize8.3cm\epsffile{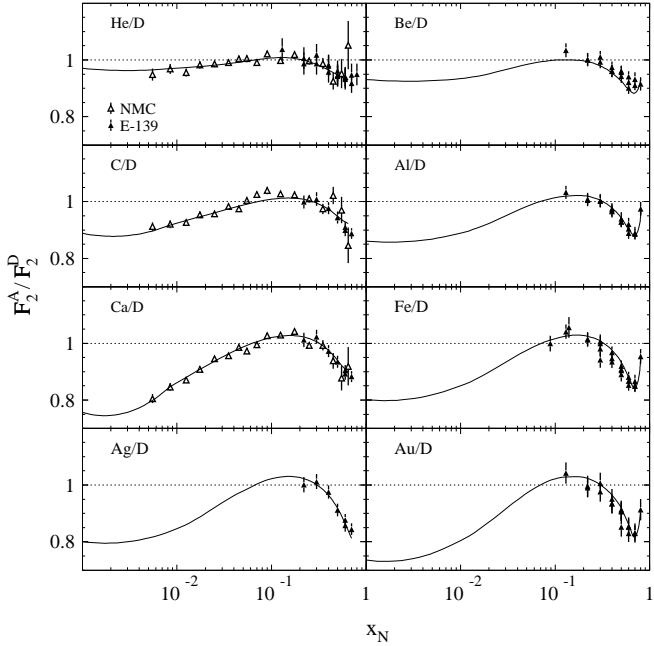}}}
\end{picture}
\caption{\label{fig:ad}{\em $F_2^A/F_2^D$ data. The lines interpolate the
values obtained with the NLO nPDF set at the respective $Q^2$, and extrapolate 
to low $x_N$ at the $Q^2$ leftmost point.
 }}
\end{figure}

The parameters were fixed minimizing the function $\chi^2$ defined in the customarily way as 
\begin{equation}
\chi^2\equiv \sum_i \frac{(\sigma^{ex}_i-\sigma^{th}_i)^2}{\Delta^{2}_i} 
\end{equation}
where $\sigma^{ex}_i$ stands the measured observable, $\sigma^{th}_i$ the corresponding NLO (LO) estimate, $\Delta_i$ the statistical and systematic errors added in quadrature, and the sum runs over all the data points $i$ included in the fit. No artificial weight has been given to any particular subset of data.

In order to compute the structure function and Drell-Yan cross-section for Deuterium, for which we neglect the nuclear corrections, and as a basis to construct the observables for the different nucleus, we use the parton distributions in the {\it free} proton as provided by the GRV98 analysis \cite{GRV98}. Consequently, we fix the initial scale to $Q_0^2=0.4$ GeV$^2$ (0.26 GeV$^2$) and the QCD scale (for five flavors) $\lambda_{QCD}^{(5)}=167$ MeV (132 MeV)  to NLO 
(LO) accuracy. The total $\chi ^2$ obtained is 300.15 units for the NLO fit and 316.35 for LO one.

The comparison between the data on the ratios of different nuclear structure 
functions to deuterium and those computed with NLO nPDF is shown in Figure 1 
(the LO prediction one is almost indistinguishable). The solid line 
corresponds to the result of the fit computed at the $Q^2$ value of each 
experimental point, whereas the extrapolation to small $x_N$ has been 
performed using the corresponding $Q^2$ of the smallest-$x_N$ point.
 For heavy nuclei, the low $x_N$ is mainly constrained by DIS ratios to carbon, as shown in Figure 2.
 The structure function ratios are useful to determine mainly the valence quark  distributions, while Drell Yan data, shown in Figure 3 becomes crucial in order to fix sea quark distributions.
The gluon distribution enters the structure function at NLO or through the scale dependence of the parton densities, making very difficult to obtain information about it in DIS experiments. The $Q^2$ dependence of $F_2^{Sn}/F_2^{C}$, shown in Figure 5, is the
most  sensitive available observable to the gluon distribution. Nevertheless, it is worth pointing out that
 there is still a large uncertainty on this  density and data from observables where the gluon distribution enters at the lowest order, like in hadronic colliders, is needed to obtain a much better constraint.

\setlength{\unitlength}{1.mm}
\begin{figure}[!]
\begin{picture}(100,87)(0,0)
\put(7,-15){\mbox{\epsfxsize8.3cm\epsffile{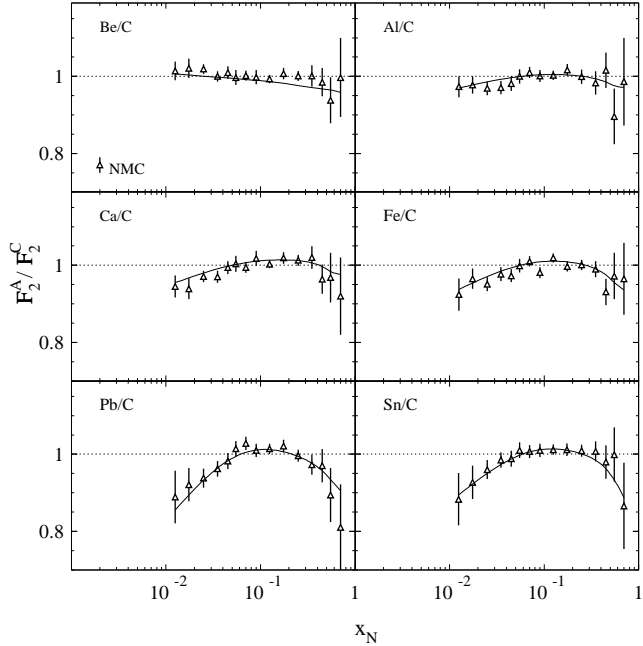}}}
\end{picture}
\caption{\label{fig:ac}{\em 
The same as Fig. 1 but for $F_2^A/F_2^C$ data
  }}
\end{figure}

The regular $A$ dependence of the parameters, as observed in Figure 5,
helps to interpolate through regions where the data is scarce and also lead 
to reasonable extrapolations where there is not available.

\setlength{\unitlength}{1.mm}
\begin{figure}[!]
\begin{picture}(100,60)(0,0)
\put(6,-47){\mbox{\epsfxsize8.5cm\epsffile{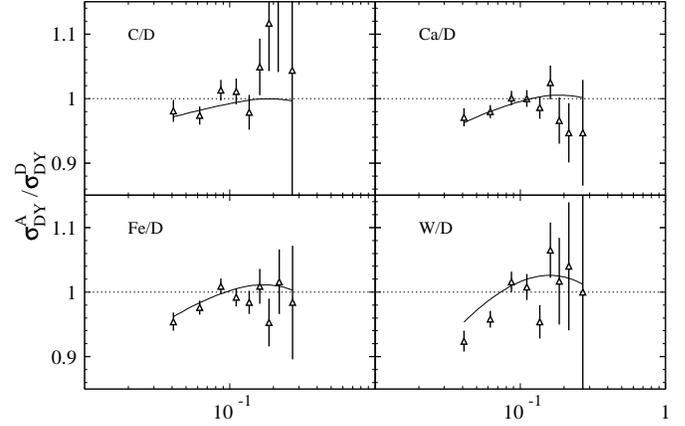}}}
\end{picture}
\caption{\label{fig:dy}{\em 
Data on nuclear Drell Yan cross sections rates to deuterium and those 
computed with NLO nPDF.  }}
\end{figure} 

\setlength{\unitlength}{1.mm}
\begin{figure}[b]
\begin{picture}(100,89)(0,0)
\put(7,-15){\mbox{\epsfxsize8.4cm\epsffile{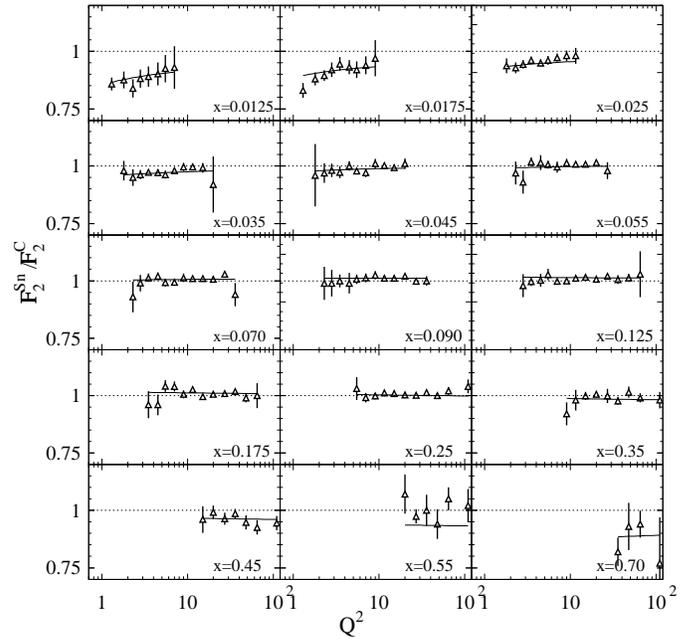}}}
\end{picture}
\caption{\label{fig:snc}{\em 
Scale dependence of $F_2^{Sn}/F_2^{C}$ data and the outcome of NLO nPDF.  }}
\end{figure} 

\setlength{\unitlength}{1.mm}
\begin{figure}[t]
\begin{picture}(100,85)(0,0)
\put(5,-20){\mbox{\epsfxsize8.3cm\epsffile{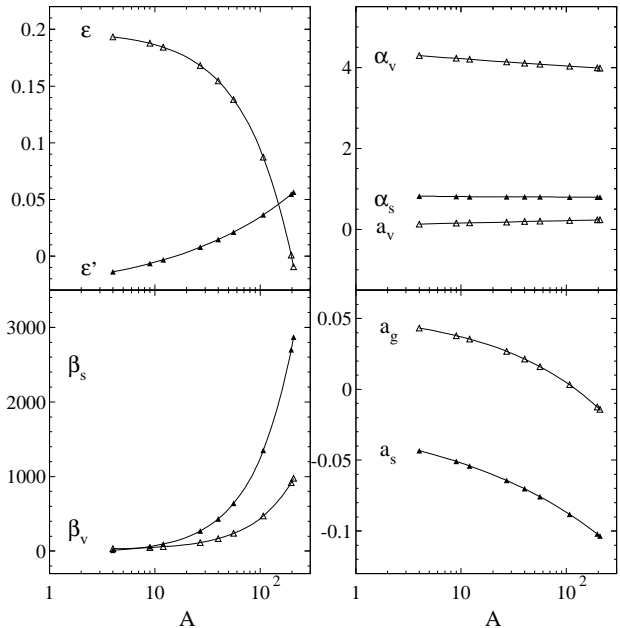}}}
\end{picture}
\caption{\label{fig:para}{\em 
A-dependence of the parameters.  }}
\end{figure} 

\setlength{\unitlength}{1.mm}
\begin{figure}[b]
\begin{picture}(100,82)(0,0)
\put(6,-19){\mbox{\epsfxsize8.5cm\epsffile{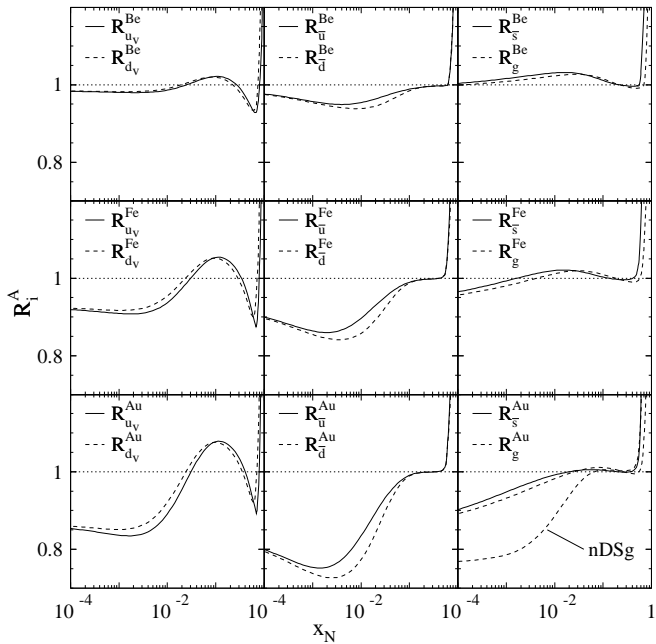}}}
\end{picture}
\caption{\label{fig:ad3}{\em 
The ratios for the densities are computed at 10 GeV$^2$
 }}
\end{figure}

Noticeably, while some parameters show a clear dependence on the size of the
nucleus, such as the shifts in the momentum fractions $\epsilon$ and 
$\epsilon'$ which drive nuclear effects at moderate and large $x_N$, those 
related to the shape of the nucleus effective densities at small $x_N$, 
such as $\alpha_v$, $\alpha_s$ and $\alpha_g=\alpha_s$ are not strongly 
dependent on $A$. The well known $A$ dependence of shadowing effects at 
small $x_N$ is driven by the normalization of these effective densities 
$a_s$, and $a_g$, and also by the large $x_N$ behavior of the densities
fixed by the parameters $\beta_v$ and $\beta_s$, which control how much of
the large $x_N$ component of the PDF enters the convolution.   

The resulting nPDF are shown in Figure 6, as ratios to free proton PDF, as 
defined in Eq.(\ref{eq:f1}), for valence quarks, light sea and strange quarks,
and gluons, at $Q^2=10$ GeV$^2$. The numerical computation of these ratios 
once the nPDF have been written and extracted within the convolution approach, 
is straightforward and allows a comparison with standard analyses, and
other parton distribution functions. 
The ratios are also provided (in a FORTRAN code) as grids in $x_N,Q^2$ and 
$A$ for practical purposes and can be obtained upon request from the authors.

Similar results are found using nucleon parton densities other than GRV98.
The similarity between modern parton densities guarantees that the nuclear 
ratios obtained with a given PDF set lead to reasonable nPDF when combined 
with another. Therefore, the parton distribution in a proton of a nucleus $A$ 
can be simply obtained by multiplying the nuclear ratios obtained in our 
analysis by any modern set of proton PDF. Of course, this is true provided 
the distributions come from analyses at the same order in QCD, as we will 
discuss in the next section.         

It is worth mentioning that the agreement between nuclear parton 
distributions and data is remarkably better in the case of convolution based 
parameterizations than the ones found with multiplicative parameterizations 
in previous (LO) analyses. Those analyses yield $\chi^2$ values around 630 in 
the case of Ref.\cite{EKS98}, and even larger values with the parameterization 
of Ref.\cite{HKM}, for the same data set used in the present analysis. 
A detailed comparison with the parameterization of Ref.\cite{EKS98} indicates
that the  $\chi^2$ of that set is basically due to a large contribution from 
the $Q^2$-dependence of $F_2^{Sn}/F_2^{C}$ data. Therefore, one might expect 
significant differences between our LO gluon nuclear ratio and that of EKS98.

In Figure 7 we show a comparison between the nuclear ratios for LO 
$u_{\rm valence}$, $\bar{u}$ and $g$ proton (in calcium) distributions from 
 nDS, EKS98 and HKM sets, at $Q^2=2.25$ GeV$^2$ and 100 GeV$^2$.

 \setlength{\unitlength}{1.mm}
\begin{figure}[!]
\begin{picture}(100,85)(0,0)
\put(8,-20){\mbox{\epsfxsize8.3cm\epsffile{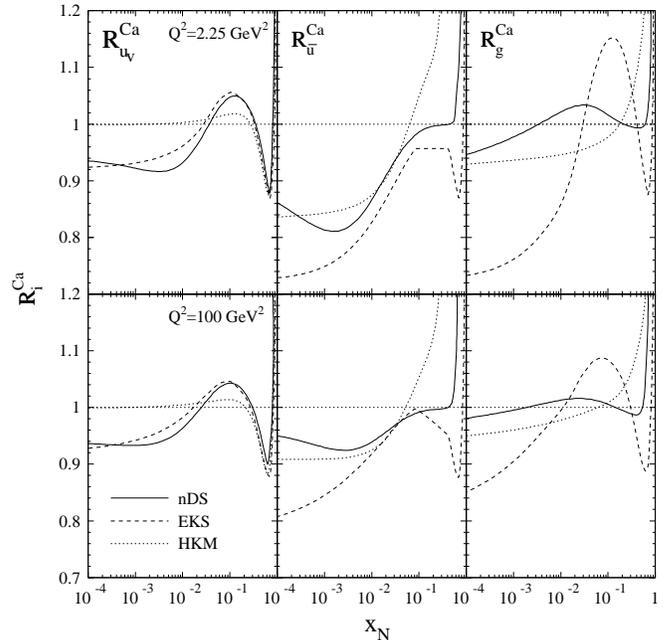}}}
\end{picture}
\caption{\label{fig:ratios1}{\em 
Ratios coming from different nPDF sets }}
\end{figure}

The main differences with previous nuclear parameterizations are  found in 
gluon densities, and to less extent in sea quarks, when comparing with
ESK98, and in valence and sea quarks in the case of HKM.   
Our gluon densities show comparatively less small $x_N$ shadowing for heavy 
nuclei than ESK98, and very little antishadowing at intermediate $x_N$, 
which is considerable in their distributions. For nuclei lighter as C our 
gluons show only a tiny antishadowing effect.

In order to study the sensitivity of different observables on the amount
of shadowing in the nuclear gluon distribution, we have performed an 
alternative extraction of nPDF from the same data set but constraining 
the gluon density in heavy nuclei to show a stronger shadowing effect at 
small $x_N$. We provide
 the result in set called nDSg, constrained to satisfy $R_g^{Au}=0.75$ at
$x_N=0.001$ and $Q^2=5$.
The $\chi^2$ value of this analysis is around $550$, considerably larger 
than the {\it unconstrained} fit, and should be considered only 
as a mean to study variations on, mainly, the gluon nuclear distribution.
An example for the gluon nuclear ratio in $Au$ is shown in the last plot in 
Figure 6.

Compared with HKM, our valence distributions show low $x_N$ shadowing, whereas
in HKM there is none. In our parameterization the ratios for sea and gluon 
densities approach unity as $x_N$ grows, while with HKM distributions these 
ratios show a strong rise. The difference in the fits may be understood due 
to the fact that both the low $x_N$ region in valence densities, and the large
$x_N$ behavior of the sea distributions have little impact in DIS observables,
and is only picked up by Drell-Yan yields, not included in HKM analysis.

\section{NLO}
\label{sec:compa}

\setlength{\unitlength}{1.mm}
\begin{figure}[t]
\begin{picture}(100,85)(0,0)
\put(7,-18){\mbox{\epsfxsize8.4cm\epsffile{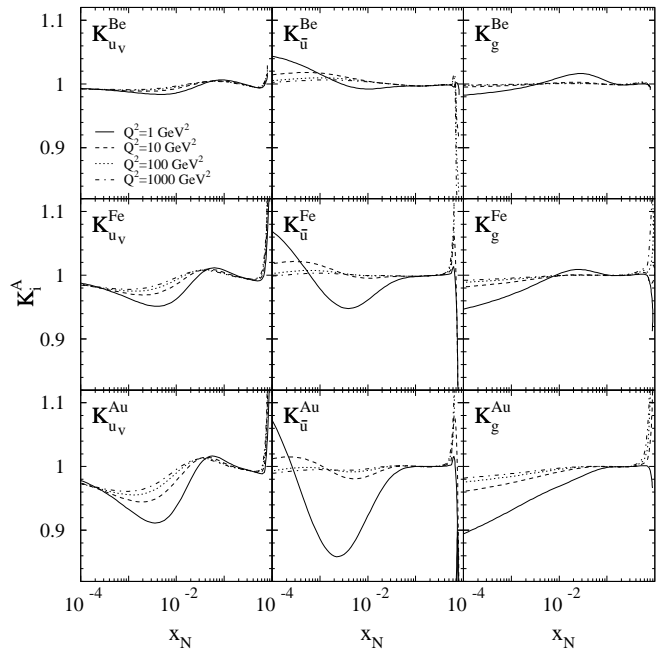}}}
\end{picture}
\caption{\label{fig:ratios2}{\em 
Ratios between NLO and LO extractions of nuclear effects for different 
nuclei.  }}
\end{figure}
Although there are no significant differences between the total $\chi^2$ 
values obtained in LO and NLO fits, the corresponding nPDF, and also the 
ratios to ordinary PDF, $R_i(x_N,Q^2,A)$, indeed differ, 
specially at low $Q^2$. In some cases, the differences are as large as the 
nuclear effects themselves. In Figure 8 the ratios between the NLO and LO 
extractions of $R(x_N,Q^2,A)$
\begin{equation}
K^{A}_i(x_N,A,Q^2)\equiv\frac{R^{NLO}_i(x_N,Q^2,A)}{R^{LO}_i(x_N,Q^2,A)}
\end{equation} 
are shown as a function of $x_N$ for various $Q^2$ and different nuclei.

The main differences between $R_i^{NLO}(x_N,Q^2,A)$ and $R_i^{LO}(x_N,Q^2,A)$
are found in sea quark and gluon densities, most noticeably at low $x_N$
and in heavy nuclei. These differences are correlated through the scale
dependence with a similar behavior in the valence ratios at small $x_N$. 
As one would expect, the differences are more significant at small $Q^2$ and 
fades away as $Q^2$ increases.

The fact that LO analyses lead to fits comparable in accuracy to NLO analyses
is due to the relatively moderated range in $Q^2$ spanned by the data, and 
the absence in the data set of nuclear observables strongly dependent on 
the gluon distributions. This, of course neither imply that the differences 
in $R(x_N,Q^2,A)$ as obtained at LO and at NLO accuracy would be negligible,
in fact beyond LO the ratios are factorization scheme dependent, nor that 
there would not be significant NLO corrections in other observables.  
This is particularly true for those that rely on the gluon or sea quark 
densities and stress the importance of having NLO extractions of nPDF.

As an example of an observable which is sensitive to NLO corrections, in 
Figure 9 we show the LO and NLO leading twist cross sections for the 
production of neutral pions in d-Au collisions as a function of the 
transverse momentum $p_T$ of the final state particle. The plots correspond 
to nucleon center of mass energies $\sqrt{s_{NN}}$ of 200 GeV and a range in 
pseudo-rapidity of $|\eta| < 0.18$, computed using the code in 
\cite{hadronicos} adapted for nuclear beams. 
Pion fragmentation functions were taken from Ref.\cite{K}.  
The LO and NLO cross section are computed at two different values for the factorization and renormalization scales $\mu_R=\mu_F= p_T$ and $\mu_R=\mu_F= 2 p_T$. The differences between both predictions give an estimate of the theoretical uncertainty in the fixed order calculation. As can be observed, there is a considerable reduction in the scale dependence of the cross section when the NLO corrections are taken into account. This feature is found in almost any infrared-safe observable in hadronic collisions, indicating that LO calculations can only provide 
a qualitative description.
The inset in Figure 9 shows the $K_{NLO}$ factor, defined as 
\begin{equation}
K_{NLO} = \f{\sigma_{dAu}^{NLO}}{\sigma_{dAu}^{LO}},
\end{equation}
i.e. the ratio between the NLO and LO cross section, computed at a given 
factorization and renormalization scales, and using the corresponding nPDFs 
and fragmentation functions. Notice that the $K_{NLO}$ factor is not a 
physical quantity, actually it turns out to be strongly scale dependent,  
but provides an estimate of the size of the NLO corrections. In this case,
it shows that the one-loop QCD corrections for pion production at RHIC are 
of the order of 50\% or even larger, while the strong increase 
in the correction when the transverse momentum  $p_T$ is smaller 
than $\sim 2$ GeV indicates we approach the limit of the region of validity 
of pQCD calculations.
\setlength{\unitlength}{1.mm}
\begin{figure}[t]
\begin{picture}(100,90)(0,0)
\put(-12,-4){\mbox{\epsfxsize10.5cm\epsffile{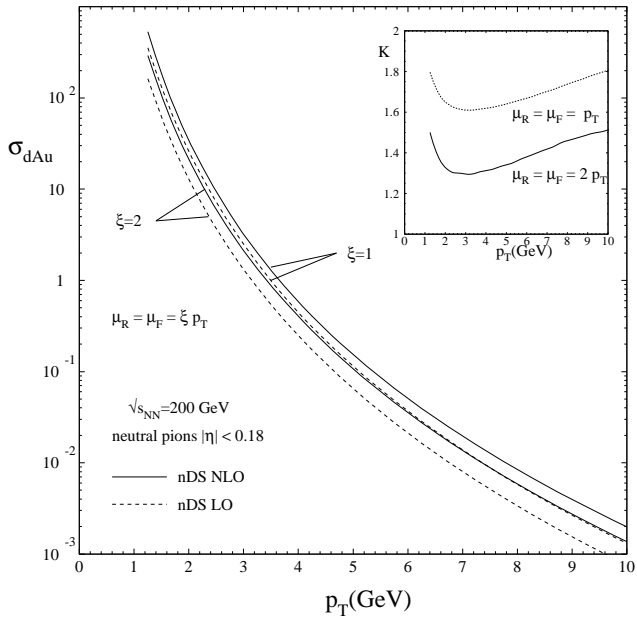}}}
\end{picture}
\caption{\label{fig:ratios3}{\em LO and NLO neutral pion production cross sections for d-Au collisions. }}
\end{figure}
\setlength{\unitlength}{1.mm}
\begin{figure}[!]
\begin{picture}(100,90)(0,0)
\put(10,-12){\mbox{\epsfxsize8.cm\epsffile{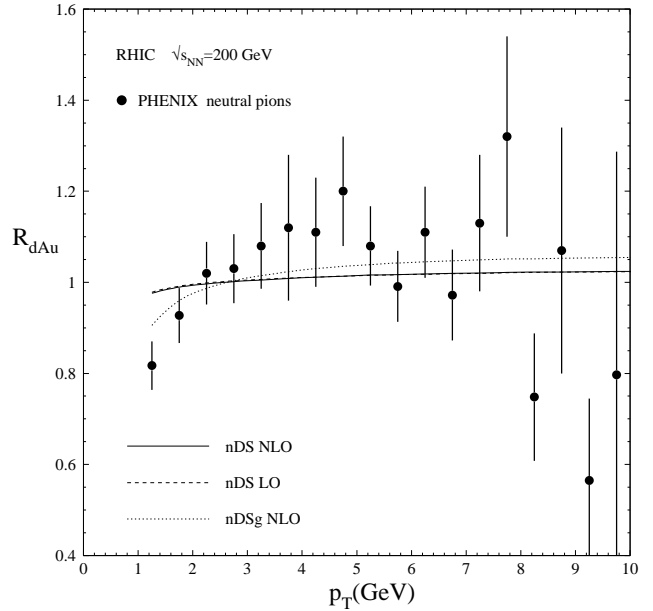}}}
\end{picture}
\caption{\label{fig:ratios4}{\em  Neutral pion production nuclear rates. }}
\end{figure} 

Finally, in Figure 10 we show the estimates for the d-Au cross sections 
to neutral pions but normalized to proton-proton cross sections, against 
the data reported by \cite{phenix}, and not included in the present fit. 
With the exception of the lowest $p_T$ data, which is in the borderline 
of the perturbative domain, the estimate agrees with the data well within 
the experimental uncertainties. Notice that even though the LO and NLO cross 
sections differ substantially, the discrepancies almost cancel in the ratios,
to proton, highlighting the perturbative stability and the consistency between
the LO and the NLO extractions of nPDF.
To quantify  the effect of nuclear shadowing in the gluon distribution on 
hadron production at RHIC we have computed the same observable
 using the alternative NLO set of nPDF (set nDSg) with a larger shadowing in the gluon distribution at small $x_N$. The Figure shows a reduction in the d-Au 
cross section at small $p_T$ for the set with larger gluon shadowing, however 
it seems unlikely to obtain a much smaller value for the ratio $R_{dAU}$ at 
$p_T\sim 2$ GeV with realistic nPDF. Notice that the cross sections receives 
contributions typically from $x_N> 0.01$. 
Similar conclusions, with a slightly bigger reduction at small $p_T$, are reached if the same observable is computed for neutral pion production in the forward region ($\eta\sim 3$). Any experimental finding on a stronger reduction of $R_{dAU}$ 
might certainly be considered as an evidence of a new phenomena, at least 
beyond  pQCD.

\section{Conclusions}

We have performed for the first time a full NLO QCD global analysis of 
nuclear DIS and Drell Yan data using a convolution approach to parameterize 
nPDF. We have found that this strategy not only leads to much more accurate 
nPDF but simplifies considerably the numerical computation of QCD corrections 
at NLO. Although both LO and NLO nPDF reproduce the available data with 
comparable precision, they show non negligible differences which have to be 
taken into account when computing other observables.

\section{Acknowledgements}

We warmly acknowledge W. Vogelsang for interesting discussions and suggestions.

\section{Appendix}

\begin{table*}
\caption{\label{tab:table3} Parameters of the NLO and LO nPDF  }
\begin{ruledtabular}
\begin{tabular}{lcccccc} 
Parameter     &   \multicolumn{3}{c}{NLO}        & \multicolumn{3}{c}{LO}      \\
  & $\gamma$ & $\lambda$ & $\delta$ & $\gamma$ & $\lambda$ & 
$\delta$  \\ \hline
$\epsilon_v$  & 0.1984 &-0.0013 & 0.0814 & 0.2030 &-0.0014 & 0.9510\\
$\epsilon'_v$ & 0.0346 &-0.0124 & 0.9421 & 0.0351 &-0.0133 & 0.3657\\
$a_v$         & 0.7546 &-0.6687 &-0.0473 & 0.7251 &-0.6647 &-0.0583\\
$\alpha_v$    & 2.1412 & 2.2690 &-0.0390 & 2.1786 & 2.5720 &-0.0439\\
$\beta_v$     &-0.0474 & 0.3730 & 1.1301 & 19.925 & 2.2760 & 1.1463 \\
$a_s$         &-0.0135 &-0.0202 & 0.2797 &-0.0179 &-0.0189 & 0.2664 \\
$\alpha_s$    & 0.7980 & 0.0814 &-0.8647 & 1.0616 & 0.0572 &-0.6277\\
$\beta_s$     &-24.325 & 7.3191 & 1.1204 &-24.107 & 7.3526 & 0.4284 \\
$a_g$         & 0.0565 &-0.0073 & 0.4244 & 0.0629 &-0.0076 & 0.4285\\ \hline
$\chi^2/d.o.f.$  &   \multicolumn{3}{c}{299.91/393}    & \multicolumn{3}{c}
{316.35/393}      \\
\end{tabular}
\end{ruledtabular}
\end{table*}

\end{document}